\documentclass[useAMS,usenatbib,usegraphicx]{mn2e}

\def\kms{$\rm km\, s^{-1}$}
\def\cm3{$\rm cm^{-3}$}
\def\Ts{$\rm T_{*}$}
\def\Vs{$\rm V_{s}$}
\def\n0{$\rm n_{0}$}
\def\B0{$\rm B_{0}$}

\def\erg{$\rm erg\, cm^{-2}\, s^{-1}$}
\def\Hb{H$\beta$}
\def\Ha{H$\alpha$}

\title[Analysis of  AG Pegasi spectra]{An analysis of the emission line spectra of AG Pegasi between phases 7.34 and 9.44}
\author[M. Contini]{Marcella Contini\thanks{E-mail:
contini@ccsg.tau.ac.il} \\
School of Physics and Astronomy, Tel Aviv University, Tel Aviv 69978, Israel\\}
\begin{document}

\date{Accepted . Received ; in original form }

\pagerange{\pageref{firstpage}--\pageref{lastpage}} \pubyear{2002}

\maketitle

\label{firstpage}

\begin{abstract}

The UV and optical spectra from the HST FOS and from the FAST spectrograph at the Whipple Observatory
reported by Kenyon et al (2001) at different phases are analysed, leading to 
new  results about the configuration of AG Peg.
The FAST spectra contain both H$\alpha$ and \Hb ~lines, whose ratio changes with phase
indicating that different nebulae contribute to each spectrum. In particular, the spectrum
emitted from  one of the nebulae must be collision dominated, in order to justify the 
relatively high \Ha/\Hb.
Consistent modelling by the code SUMA, which accounts  for the coupled effect of  the photoionisation
from the hot star and the shock, shows  the important role of at least three nebulae:
1) the nebula between the stars, downstream of the  shock created by collision of the winds,
which propagates in reverse towards the WD,
2) the nebula downstream of the shock expanding in the outskirts of the giant atmosphere, and 
3) the shock propagating outwards the binary system, that is not  reached by  the hot source radiation.
The relative contribution of the  three nebulae to the final spectra at
different phases are calculated for all the lines.
It is  found  that the contribution to the different lines changes with the viewing 
angle of the system.
The relatively low preshock density and magnetic field adopted in the  modelling of
the expanding shocks reveal that they are
merging  with the ISM, thus  explaining the slow decline of AG Peg.

\end{abstract}

\section{Introduction}

AG Pegasi (AG Peg) is a symbiotic   binary consisting of a white dwarf (WD) and
a  red giant  (spectral type M3).
A continuing slow decline of the far-UV and near-UV continuum is noticed by
Kenyon, Proga, \& Keyes (2001). They claim that the eruption 
of the hot component in AG Peg may be the slowest classical nova outburst ever recorded.
AG Peg evolution   was followed by the 
observations (Kenyon et al 1993, Kenyon et al. 2001) and  modeled by the interpretation of the
line and continuum spectra (e.g. Contini 1997).

The collision of the  winds from the hot star  and  the red giant  in symbiotic binaries
 creates a complex hydrodynamic structure (Nussbaumer 2000 and references therein).
The  emitting gas within the system is then ionised and heated both by
 the photoionising flux from the hot star and by shocks.
In a previous paper (Contini 1997, hereafter Paper I) the observed continuum and line spectra
from AG Peg  at phases  earlier than $\phi$ = 7.12, have been explained by composite models
accounting both for  photoionisation and shocks. Indeed, the complex structure
of the system requires  schematic models which  represent roughly
the real picture. The consistency of calculation results for the line
spectra in the UV and in the optical ranges,  as well as for the spectral energy
distribution (SED) of the continuum leads to a better  understanding of the
physical characteristics of the system.

Following the results of Paper I, two shock fronts are created at collision
of the winds between the stars,  one propagating in reverse 
toward the WD and  the other propagating outwards the system.
Moreover, recent observations of symbiotic novae show that spectra are also
emitted from  downstream of shocks propagating outside the system and eventually
merging with the ISM (Contini \& Formiggini 2001).
At phase 7.05  of AG Peg the expanding shock has already passed over the  red giant
and slightly accelerates propagating in the atmosphere of the red giant, 
opposite to the WD, where the density gradient is decreasing.

New observations are presented by Kenyon et al. (2001, hereafter KPK01). They provide
optical and UV spectra at phases 7.34 - 9.44, acquired with Faint Object Spectrograph (FOS) on board of
 the Hubble Space Telescope (HST) between late 1993 and late 1996. 
Low resolution optical spectra with the FAST spectrograph mounted at the F.I. Whipple Observatory 
1.5-m telescope are also reported.
After an accurate analysis of the data, KPK01 conclude that
"if the hot component emits enough high 
energy photons to account for the forbidden line fluxes, mechanical heating
from the colliding winds may explain the high electron temperature.  {\it A detailed
photoionisation calculation which includes shock excitation would test this
proposal.}"  
Responding to  KPK01, I   present further modelling of AG Peg on the basis of their observational
data at the latest phases ($\phi$ = 7.39 - 9.44), adopting for  the calculations
the code SUMA (Viegas \& Contini 1994, Paper I). 
SUMA calculates the emitted spectrum from the gas, accounting for the coupled effect
of  shock and photoionisation from an external source.
The spectra in the different phases are  considered in Sect. 2.
The models  and the modelling  process are presented in Sects. 3 and 4, respectively. 
Results are discussed in Sect. 5 and concluding remarks appear Sect. 6.

\section{The spectra at different phases}

\begin{figure}
\includegraphics[width=84mm]{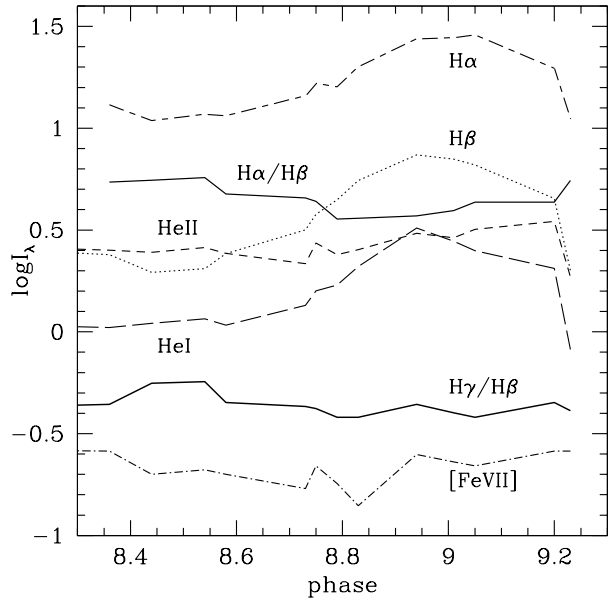}
\caption
{The evolution of some significant line intensities with phase.
The lines are HeI 4471, HeII 4686, [FeVII] 6087, \Hb, and \Ha.}
\end{figure}

Line fluxes observed by KPK01 were corrected for E(B-V)=0.1.
There are several values for  E(B-V)  in the range 0.08-0.15
obtained by different methods
(Tomov, Tomova, \& Raikova 1998, Penston \& Allen 1985, etc.).
Some significant line  fluxes  and line ratios corrected for E(B-V) = 0.1 are given in Fig. 1 as function
of the phase.
Fig. 1 shows that \Ha/\Hb   ~and H$\gamma$ /\Hb ~at  phases  $\phi$ $<$  8.8
are higher than  case B values.
A higher reddening correction  of the spectra is not predicted because
it would lead to  very high values of the far UV line fluxes.
Moreover, KPK01 notice that the depth of the 2200 \AA
~absorption feature does not vary with phase and is consistent with E(B-V) = 0.1.
Finally, the best fit of the calculated  continuum SED
to the data of the IR dust bump at early phases (Paper I) corresponds to  a relatively
low gas-to-dust ratio, d/g = 2. 10$^{-15}$ by number.

We would like to investigate  the origin of such high values of \Ha/\Hb ~and H$\gamma$/\Hb. 
We will refer to radiative transfer effects of HI,
to collision excitation of H lines, and, more specifically, to
 the effect of the viewing angle of the emitting nebulae.

Recall that densities in the emitting nebulae  of symbiotic systems are generally
high ($>$ 10$^4$ \cm3), sufficiently high that
some self absorption occurs in the Balmer lines. This leads to
a marked strengthening of the \Ha ~line relative to the rest of the Balmer series.
In case B, as  the optical thickness of Ly$\alpha$ ~increases, the first effect in the decrement
occurs when \Hb ~converts into \Ha ~ (Cox \& Mathews 1969).
Particularly, for large optical thickness of the Ly$\alpha$ ~line, 
the increase of the \Ha ~optical thickness leads to an increase 
of both  \Ha/\Hb ~and H$\gamma$/\Hb ~(Cox \& Mathews 1969, Fig. 1).
In fact,  \Ha ~is mostly scattered because any L$\beta$ photons it forms are quickly 
absorbed and converted back to \Ha, while  \Hb ~is absorbed and converts to \Ha + P$\alpha$.

Moreover, Drake \& Ulrich (1980) claim that
\Ha/\Hb ~ratios greater than 10 can be obtained with optical depths of the order of 100 or more by Balmer 
self-absorption alone.
Particularly, conditions which tend to produce very steep Balmer decrements are found to be
large optical depths in \Ha, low photoionisation rates, and values of the electron density 
between 10$^8$ \cm3 and 10$^{12}$ \cm3
(see also Krolik  \& McKee 1978).

In the case of symbiotic systems, collisional processes are very effective,
because shock fronts are created by collision of the winds between the stars
with velocities $\geq$ 40 \kms. Such velocities correspond 
to temperatures  $\geq$ 20,000 K in the postshock region.
Higher velocities are actually  observed leading to higher temperatures.
Collisional excitation increases with temperature and prevails for T $\geq$ 20,000 K.
Collision strengths are $\propto$ E$_{ij}^{-2}$ exp(- E$_{ij}$/kT) (Contini \& Aldrovandi 1983)
where E$_{ij}$ is the energy of the level.
Radiation from the hot star, on the other hand,  cannot heat the gas to $\geq$ 20,000 K.
So, high  \Ha/\Hb ~line ratios are predicted  in the spectra emitted
from  shock dominated nebulae.

It was  shown in  Paper I that the  spectra  of AG Peg at different epochs are
emitted  mainly by  two nebulae : one in the downstream region
of the shock front propagating in reverse towards the hot source and the other
downstream of  the shock front  expanding outwards in the atmosphere
of the red giant opposite to the hot source.
In particular, the gas is collisionally heated in the outer  region of the expanding nebula,
while  is heated by radiation in the internal region of the same nebula facing the hot source. 
So,  collision dominated spectra will prevail at photometric minimum.
Fig. 1 shows that the \Ha/\Hb ~ line ratio has a rough maximum at phase $\sim$ 8.5, 
which could be due to collisional effects.
We will come back to this point after modelling the spectra.

 When the system evolves through
the photometric minimum  and the hot source is occulted, the HeII lines that are 
emitted   by a  photoionised gas, should strongly decline. 
On the contrary,  Fig. 1  shows that the  HeII 4686 flux does not change  with phase 
more than about 22\% throughout the period between phases 8.33 and 9.44. 
This behavior
reveals  that  nebulae heated and ionised by  radiation   contribute to the spectra   near
maximum as well as near photometric minimum.

\section{The models}

SUMA calculates the emitted spectrum from a gas in a  plane parallel geometry.
The  nebula between the stars
is ionised and heated by both black body radiation from the WD and  the (reverse) shock.
Therefore, radiation reaches the very   front edge  of the shocked  nebula.
On the other hand, radiation and shock act on opposite edges of the nebula
downstream of the expanding shock which propagates in the red giant atmosphere.

The input parameters are : the shock velocity, \Vs, the preshock density, \n0,
the preshock magnetic field, \B0, the colour temperature of the hot star,
\Ts, the ionisation parameter, U, the geometrical thickness of the emitting
gaseous nebula, D,  and the abundance of He, C, N, O, Ne, Mg, Si, S, A, and Fe relative to H.
For all models  \Ts=10$^{5}$ K. 

We calculated a large grid of models with the following criteria:
to fit the most significant line ratios  and to be consistent with the modelling reported in
Paper I which was  for earlier photometric phases.

Three models were selected from the grid, m1, m2, and m3. 
Model m1 represents the nebula in the downstream region
of the  shock propagating in reverse toward the WD. The downstream nebula, facing the red giant,
is however close to the WD and the photoinising flux from the hot source  reaches the shock front edge
of the nebula. Model m2 represents the nebula downstream of the shock
propagating outwards the system in the red
giant atmosphere opposite to the WD. The downstream nebula toward the WD is ionised and heated 
by both the shock and radiation from the hot source. 
Model m3 shows the same shock characteristics
of model m2, but the emitting nebula downstream is not reached by the  WD radiation flux.
Model m2 is radiation dominated (RD), while m3 is shock dominated (SD).
Model m3  being SD, the  emitted spectrum is  mainly  excited by collision.
Recall that at later phases the shock front, merging with the ISM, is disrupted by 
 R-T and K-H instabilities (see Contini \& Formiggini 2001).
Therefore, the ionising flux from the hot source will be eventually prevented from reaching
the nebula in the downstream region of the shock  by intervening  matter.

So,  in the following, the "reverse shock" will refer to model m1, the "RD expanding shock" to model m2, 
and the "SD expanding  shock" to model m3.
Models m1 and m2 are matter-bound, while m3 is radiation bound (see Sect. 4.2).

\section{Modelling the spectra}

The spectra contain lines in a relatively large range of ionisation levels
and in the  different wavelength domains.
Recall that   the distribution of the fractional abundances
of the ions   downstream (Fig. 2) depends on the input parameters and that the line intensities
  increase with the volume of gas  emitting the
corresponding ion and  with the square of the  density.

The choice of the models depends  on line ratios from  different ionisation levels 
 and is constrained by the general model of the AG Peg system.

Between  $\phi$ = 7.05 and $\phi$ = 8.07  the conditions of the reverse shock (between the two stars)
could not change sensibly because the reverse shock is most likely a standing shock,
so, we adopt  input parameters  at $\phi$ = 8.07,   \Vs=60 \kms and  \n0=10$^6$ \cm3
similar to those found at $\phi$ = 7.05,  \Vs=40 \kms and \n0= 3 10$^6$ \cm3 ( Paper I, Table 4, column 2).
In particular, \B0 = 10$^{-3}$ gauss is adopted for m1, which is suitable for isolated giants
(Bohigas et al 1989).

 The preshock density of the  shock   propagating throughout the atmosphere of the
red giant in the present modelling (10$^3$ \cm3) is lower
than that calculated at phases 7.05 - 7.12 (Paper I, Table 4) by a large factor ($\sim$ 3000).
The shock velocity is  slightly lower (100 \kms at $\phi$=8.07 and 150 \kms at $\phi$ = 7.05). 
A lower \n0 indicates that the shock has proceeded towards the outskirts of the giant
atmosphere. In about 820 days (between $\phi$ =7.05 and $\phi$ =8.07) the  shock expanding
with a velocity of $\sim$ 100 \kms could reach a distance of $\sim$ 7. 10$^{14}$ cm and 
is propagating  outwards  the system, merging with the ISM. Therefore, the preshock density
of model m2 is low and the magnetic field is also reduced to 10$^{-4}$ gauss. Both \n0 and \B0 are, nevertheless, 
still higher than  found in the ISM (1-10 \cm3, and 10$^{-5}$ - 10$^{-6}$ gauss, respectively).

The relative abundances adopted at $\phi$ $\geq$ 8.07 are
C/H = 3.3 10$^{-4}$, N/H = 5.3 10$^{-4}$, O/H = 9.6 10$^{-4}$, Ne/H = 8.3 10$^{-5}$, Mg/H = 2. 10$^{-5}$, 
Si/H = 3.3 10$^{-5}$, S/H = 1.6 10$^{-5}$, A/H = 6.3 10$^{-6}$, and Fe/H = 4. 10$^{-5}$, as were found
at $\phi$=7.05-7.12  (Paper I, Table 4).
Mg/H is, however, similar to that at $\phi$ = 2.74 (Paper I, Table 1) because at $\phi$ = 7.05
Mg lines were not observed and the Mg/H abundance ratio  could not be   determined by modelling the 
corresponding lines.
The adopted relative abundances 
show   that N/H and O/H ratios  are  higher than solar.
Relatively high values of N/H and O/H  are characteristic of symbiotic systems.

The modelling process of the spectra is  roughly described in the following.  Recall that model m1
represents the reverse shock, m2  the RD expanding shock , and m3 the SD expanding shock.
The results corresponding to  single models are  summed up  adopting relative weights
at each phase and the final results are compared  with the observations.
The relative weights
which lead to the best fit of the observed  spectra in the different epochs  are related to the
configuration of the system  and are discussed in the next section.

High ionisation-level lines  (e.g. NV, [MgV], [FeVII], etc) increase   with the ionisation parameter  U and with \Vs.
In fact, the temperature downstream behind the shock front is $\propto$ \Vs $^2$. 
NV/CIV ratios calculated by model m1 with relatively low velocities (\Vs$<$ 80 \kms) 
 are  lower than observed,  therefore, they must be  counter-balanced
by  line ratios higher than observed, calculated by  the  m2  model.  
The NV/CIV ratio calculated by SD models (m3)
with \Vs $\sim$ 100 \kms  is very low. So, the only way to obtain a high NV/CIV is to
adopt a high U in model m2. This  leads  to high   HeII/CIV line ratios.

The  NV/CIV ratios  calculated by a RD model with ionisation and shock acting on
opposite edges, are not high enough, if the density is high (10$^6$ \cm3). Also,  at high densities, 
the ratio of the intermediate ionisation-level 
lines (NIII/CIV, SiIII/CIV, and CIII]/CIV) would be higher than observed by a factor $>$ 10.
High densities are thus rejected for model m2. This is consistent
with  the previous discussion  (Sect. 3), namely,  that the RD expanding  
shock propagates in a low density medium.

Low ionisation-level line ratios (e.g. CII]/CIV and MgII/CIV) are low for models m1 and m2 (Table 1).
Low ionisation-level lines are high in SD models,
therefore, model m3 must be accounted for. The relatively high  CII]/CIV and MgII/CIV  ratios  
calculated by m3 counter-balance  in the averaged spectra
 the low values found by models m1 and m2.

A large grid of models is calculated in order to achieve the best fit of all
the line ratios. 

\subsection{Calculated line ratios}

\begin{table}
\centerline{Table 1}
\centerline{Line ratios to CIV 1550 from selected models}           
\begin{tabular}{ lllll} \\ \hline\\
\ line     & m1   & m2  & m3   \\ 
\ NV 1240 &     0.70   & 3.22&  0.1 \\
\ OI 1305 &    0.0   & 0.0 &  1.2(-3)\\
\ SiIV] 1394   &  0.04 & 0.04 & 0.19  \\
\ SiIV, OIV] 1403  &  0.35   &1. & 0.23  \\
\ NIV] 1486 &    0.8    & 1.17    &  0.37  \\
\ CIV 1550  &        1    &1 &   1  \\
\ HeII 1640 &     0.63    & 2.44    &  2.4(-3) \\
\ OIII] 1664    &  0.025   & 0.012    &  0.24  \\
\ NIV 1719 &    6.(-3)    &  0.04    & 5.(-6) \\
\ NIII] 1750 &  0.08       & 0.04    &  0.85 \\
\ SiIII] 1892 &    0.006    & 3.(-3)   &  0.26  \\
\ CIII] 1908 &     0.032   & 0.013&     0.76     \\
\ CII] 2325 &     5.(-5)  & 8.(-6)   &  0.35 \\
\ [MgV] 2783 &     0.024  & 0.15    &  3.(-5) \\
\ MgII 2793 &     1.3(-4) &2.(-5)& 0.046\\
\ [NeV] 2976 &    6.5(-4) & 3.(-3) &  6.(-6)\\
\ HeII 3203 &    0.038    & 0.15     &  1.4(-4) \\
\ [NeV] 3426 &   0.1      & 1.0   &  1.5(-4) \\
\ HI  4340 &     0.04    & 0.145     &  4.(-3)\\
\ HeI 4471 &     6.2(-5) & 7.(-5)    &  6.(-4) \\
\ HeII 4686 &    0.087   & 0.34    &  2.(-4)\\
\ [FeVII] 6087 & 0.045 &0.44 &2.2(-5)\\
\ CIV/\Hb      & 12.4  & 3.22 & 84.4 \\
\ \Ha/\Hb      & 2.78  & 2.78  & 7. \\
\ \Hb (\erg)& 237. & 0.003  &   4.2(-4) \\
\hline\\
\   \Vs (\kms) & 60.   & 100.   & 100. \\
\   \n0 (\cm3) & 1(6)  & 1(3)   & 1(3) \\
\   \B0 (10$^{-4}$ gauss)& 10. & 1. & 1. \\
\   \Ts (10$^5$ K)      & 1.   & 1.  & 1. \\
\   U                  &  0.5  & 5.  & - \\
\   D (10$^{14}$ cm) &   1.5   & 2.  & 2. \\
\   s                &   0    & 1    & 1   \\
\hline\\

\end{tabular}

\end{table}

The line ratios calculated by models m1, m2, and m3 are presented in Table 1.
The input parameters are given in the bottom of  the table.
The parameter s, which appears in the last row of Table 1 distinguishes between the
two cases:  radiation from the star reaches the shocked edge
of the emitting nebula (s=0) and radiation and shock act on opposite edges (s=1).
The calculated absolute \Hb ~flux  and the CIV/\Hb ~line ratios are also given
in order to  calculate the flux
of  the lines. Recall, however,  that the spectra are calculated at the nebula
but    are observed at Earth. The adjusting factor depends on the distance to Earth of
AG Peg, as well as on the distances of the nebulae  to the WD.

\subsection{Distribution of the different ions throughout  the nebulae}

\begin{figure}
\includegraphics[width=78mm]{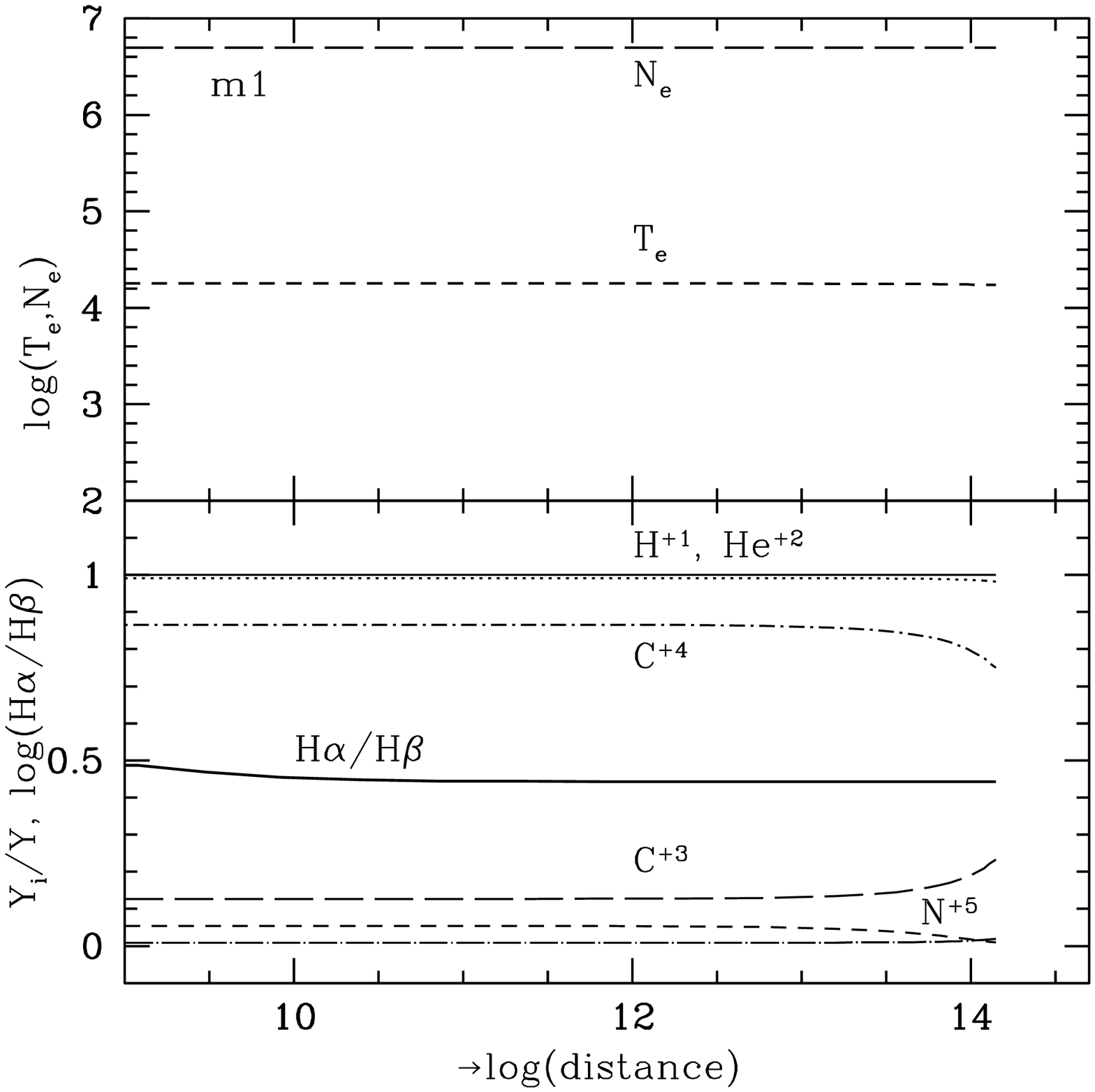}
\includegraphics[width=78mm]{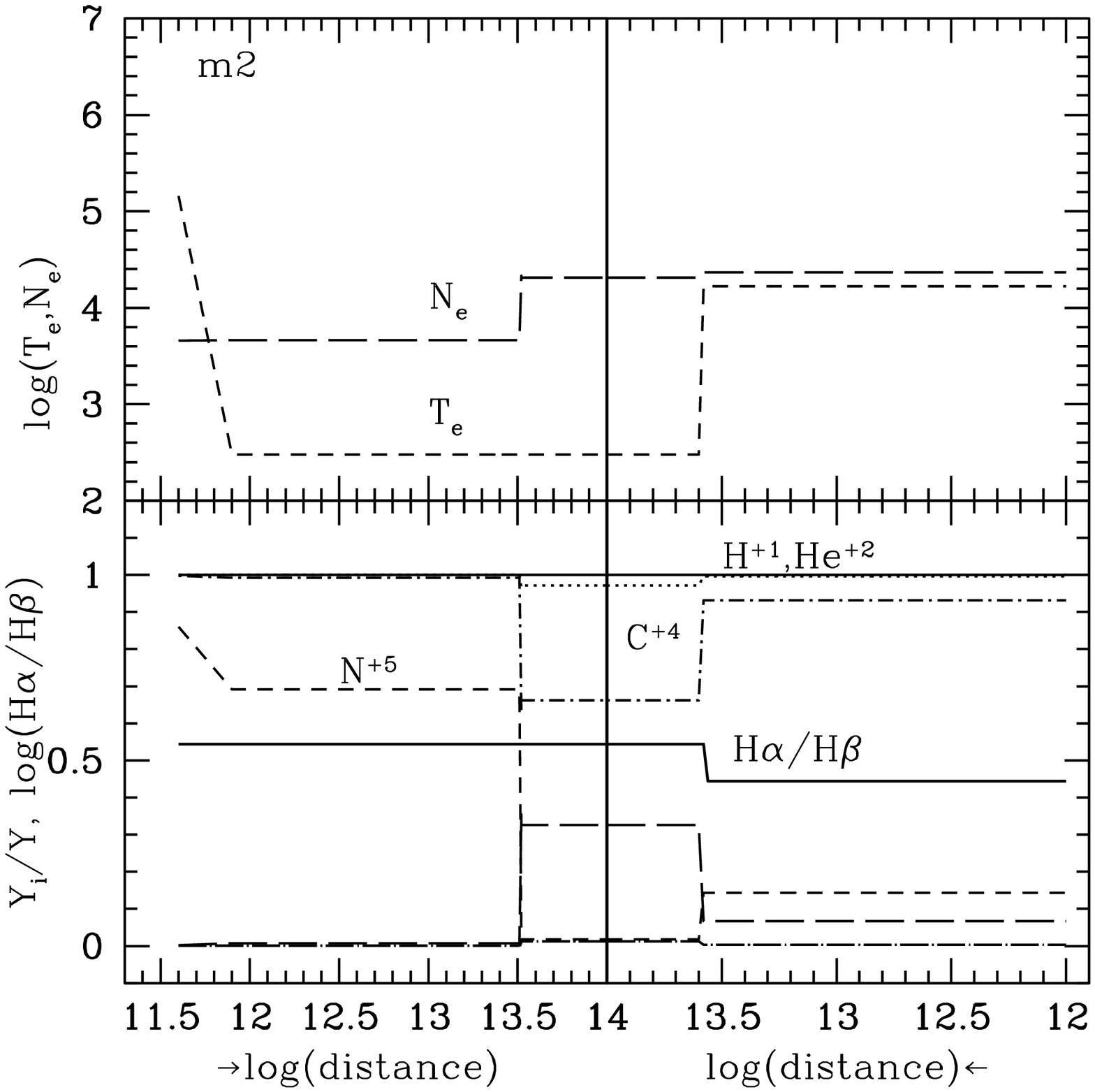}
\includegraphics[width=78mm]{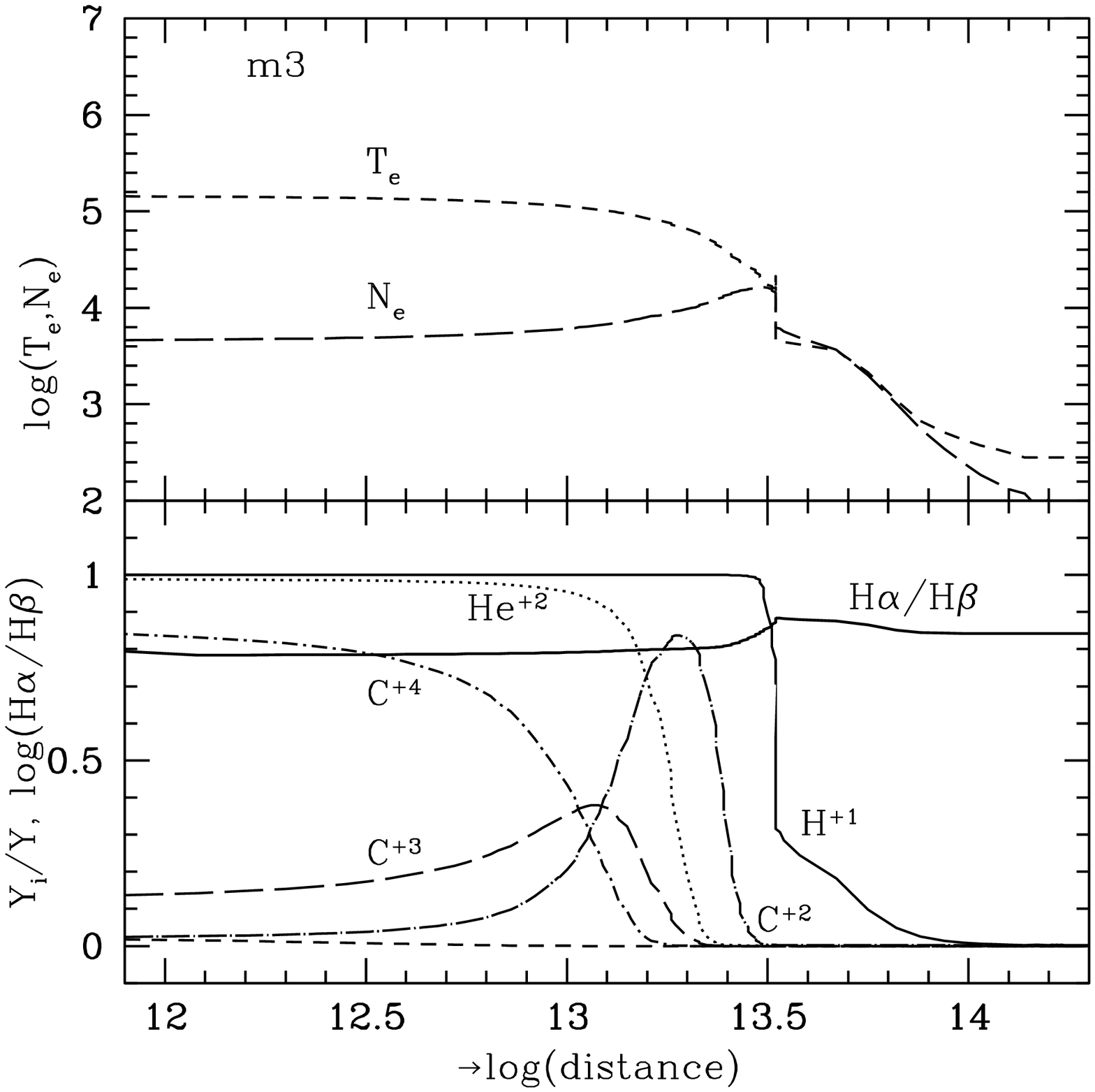}
\caption
{see text}
\end{figure}

In order to better understand the line ratios calculated by the different models, 
the distribution of the  fractional abundance of some interesting ions
as well as of the electron density and  temperature downstreams is shown in Fig. 2 
for models m1 (top diagram), m2 (middle diagram), and m3 (bottom diagram).
The shock front is on the left; black body
radiation reaches the left edge in the top diagram (m1) and
the right edge in the middle diagram (m2).
The middle diagram is divided in  two equal symmetric parts  in order to
give equal importance to  the shock dominated region (left) and to the
radiation dominated region (right).
Model m3 (bottom diagram) is shock dominated (U=0.).
In all the diagrams the ions  are represented  by the following symbols.
Solid lines : H$^{+1}$; dotted lines : He$^{+2}$; short-dashed  lines:
N$^{+5}$; dash-dotted lines : C$^{+4}$; long-dashed lines : C$^{+3}$;
long-dash-dotted lines : C$^{+2}$.

Fig. 2 shows that the temperature throughout the nebulae never
decreases below 10$^4$ K downstream of the  reverse shock (top diagram). 
This is due to the coupled effect of a relatively high U and  a relatively small geometrical 
thickness of the nebula, too small to  host a large recombination region.

The profile of the temperature is different in the RD expanding nebula (middle diagram).
The gas behind the shock front is heated to T $>$ 1.5 10$^5$ K by the shock corresponding to \Vs = 100 \kms.
The high ionisation level lines (e.g. NV) are strong due to the coupled effects of the relatively high
temperature and the strong radiation flux from the hot source.
The cooling rate ($\propto$ n$^2$) of the gas  is  sped up by free-free, free-bound, and line emission,
therefore a large cool region appears in the left side of the diagram.
Notice that  H remains  fully ionised and the
 N$^{+5}$/N fractional abundance is high in the shocked side of the nebula, even if the temperature drops,
because they are maintained by the strong radiation flux from the hot source.
C$^{+4}$ prevails throughout the nebula corresponding to the m1 and m2 models owing to the high U.

The shock dominated model (m3) shows  different conditions, characteristic of SD models. 
A relative large region is dominated by the C$^{+2}$ ion.
The  absolute flux of the \Hb ~line is  low because the emitting region throughout the nebula
is relatively small.

Interestingly, \Ha/\Hb ~ is higher in the postshock region (left side of the middle diagram) of model m2
than in the region reached by the radiation from the hot source (right side) and is definitively high
throughout all the nebula represented by model m3.

\section{Results}

The best  modelling of AG Peg should,  indeed, imply a huge
number of models in order to represent the complex nature of symbiotic systems
(see Nussbaumer 2000). The models, however, are schematized  to  three main types
representing the three emitting nebulae described in the previous sections;
therefore, the results  show only a consistent approach.

\subsection{Comparison of calculated with observed line ratios}

\begin{table*}
\centerline{Table 2}
\centerline{The comparison of calculated and observed line ratios to CIV 1550}           
\begin{tabular}{ llllllllll ll   } \\ \hline\\
& \multicolumn{2}{c}{$\phi$=8.07}&
  \multicolumn{2}{c}{$\phi$=8.33}&
  \multicolumn{2}{c}{$\phi$=9.27}&
  \multicolumn{2}{c}{$\phi$=9.44}\\
\ line     & obs  & calc       & obs    & calc    & obs  & calc   & obs  & calc \\ 
&&&&&&&&&&\\ \hline
\ NV 1240 &     0.90   & 0.8 &  1.16   & 1.3  &      1.04   & 1.18 &     1.19   & 2.39 \\
\ OI 1305 &    0.06   & 0.0001 & 0.031  & 0.0002 &     0.044  & 0.0001 &     0.03   & 0.0003\\
\ SiIV] 1394   &  0.06 & 0.05 & 0.05   & 0.06 &    0.054  & 0.057 &     0.04   & 0.08 \\
\ SiIV, OIV] 1403  &  0.22    &0.35 & 0.56  &0.40  &  0.317   &0.38   & 0.50    &0.47 \\
\ NIV] 1486 &    0.21   & 0.8     &  0.27   & 0.84   &  0.262  & 0.83  &  0.47   & 0.98 \\
\ CIV 1550  &        1    &1. &   1  &       1  &1. &   1  &       1  &1.\\
\ HeII 1640 &     0.61    & 0.67    &  1.33    & 1.1     &  1.276  & 1.0   &  2.58    & 1.90 \\
\ OIII] 1664    &  0.18    & 0.03     &  0.25    & 0.05    &  0.227 &   0.05  &  0.41    & 0.07 \\
\ NIV 1719 &     0.05    & 0.01     &  0.10    & 0.014    &  0.085  & 0.012  &  0.12    & 0.03 \\
\ NIII] 1750 &     0.05    & 0.057   &  0.09    & 0.10       &  0.084   & 0.085     &  0.14    & 0.13 \\
\ SiIII] 1892 &     0.05    & 0.02     &  0.06    & 0.05    &  0.056   & 0.04      &  0.08    & 0.07 \\
\ CIII] 1908 &     0.05    & 0.06 &     0.10    & 0.14     &  0.093   & 0.11      &  0.17    & 0.20 \\
\ CII] 2325 &     0.01    & 0.016    &  0.007   & 0.06  &     0.01    & 0.04     &  0.04    & 0.09 \\
\ [MgV] 2783 &     0.012   & 0.027    &  0.038   & 0.05      &  0.038   & 0.048    &  0.08    & 0.11\\
\ MgII 2793 &     0.02   &0.002&   0.024      & 0.01  &   0.022   & 0.013 &   0.03    & 0.014\\
\ [NeV] 2976 &    0.004   & 0.001  &  0.006   & 0.0012&  0.005   & 0.001 &     0.01    & 0.002\\
\ HeII 3203 &     0.04    & 0.04      &  0.103   & 0.07    &  0.01    & 0.05      &  0.21    & 0.11 \\
\ [NeV] 3426 &    0.022   & 0.13  &  0.073   & 0.35    &  0.023   & 0.30      &  0.15    & 0.74 \\
\ HI  4340 &     0.056   & 0.04      &  0.07    & 0.06    &  0.048   & 0.058    &  0.113   & 0.110\\
\ HeI 4471 &     0.004   & 0.0001    &  0.007   & 0.0001    &  0.005   & 0.0001     &  0.010   & 0.0002\\
\ HeII 4686 &    0.06    & 0.08    &  0.159   & 0.14    &  0.131   & 0.129    &  0.273   & 0.250\\
\ [FeVII] 6087 $^1$  & - & 0.06 &0.019&0.15  &0.012& 0.128&-&  - &\\
\  &&&&&&&   \\
\hline
\   w(m1)  &         - &      1.7(-3) &        -&   1.3(-4)&   -&  2.(-4)&    -&  0.0\\
\   w(m2)  &         - &     20. &        -&   20.&   -& 20.&     -& 20. \\
\   w(m3) &        - &      6.67&        -& 3.0 &   -&3.0 &      -&   2. \\ 
\hline
\end{tabular}

\flushleft

$^1$ [FeVII] flux from the FAST spectra at $\phi$=8.36 and $\phi$=9.27

\end{table*}

The calculated spectra at each phase  are  compared with observations in Table 2. Each
spectrum results from the weighted sum of the three models.
The  relative weights of the models  are shown in the last rows of Table 2.
The weights are very different for different models because
they depend on the emitting area of the corresponding nebula.

\begin{figure}
\includegraphics[width=84mm]{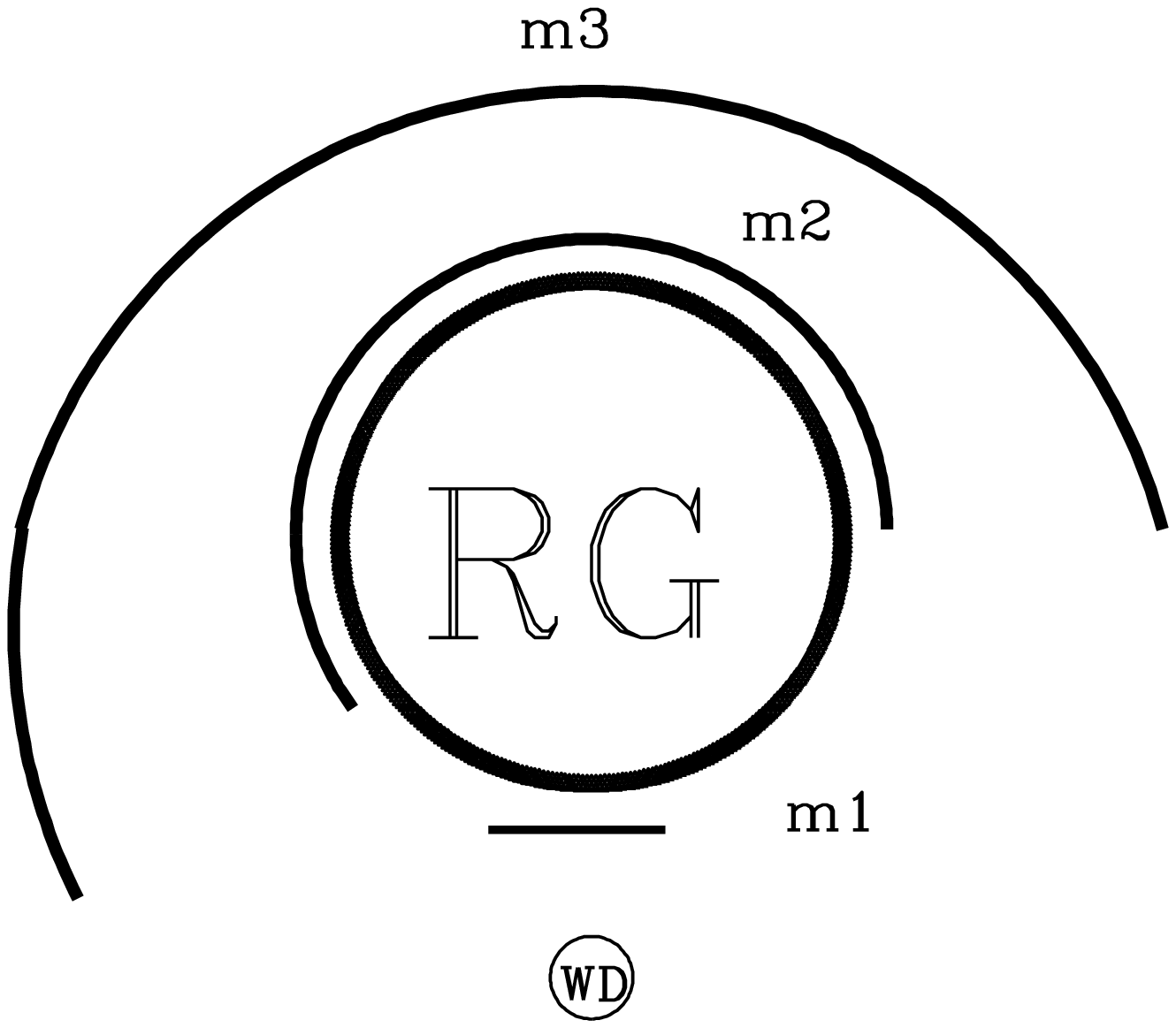}
\includegraphics[width=84mm]{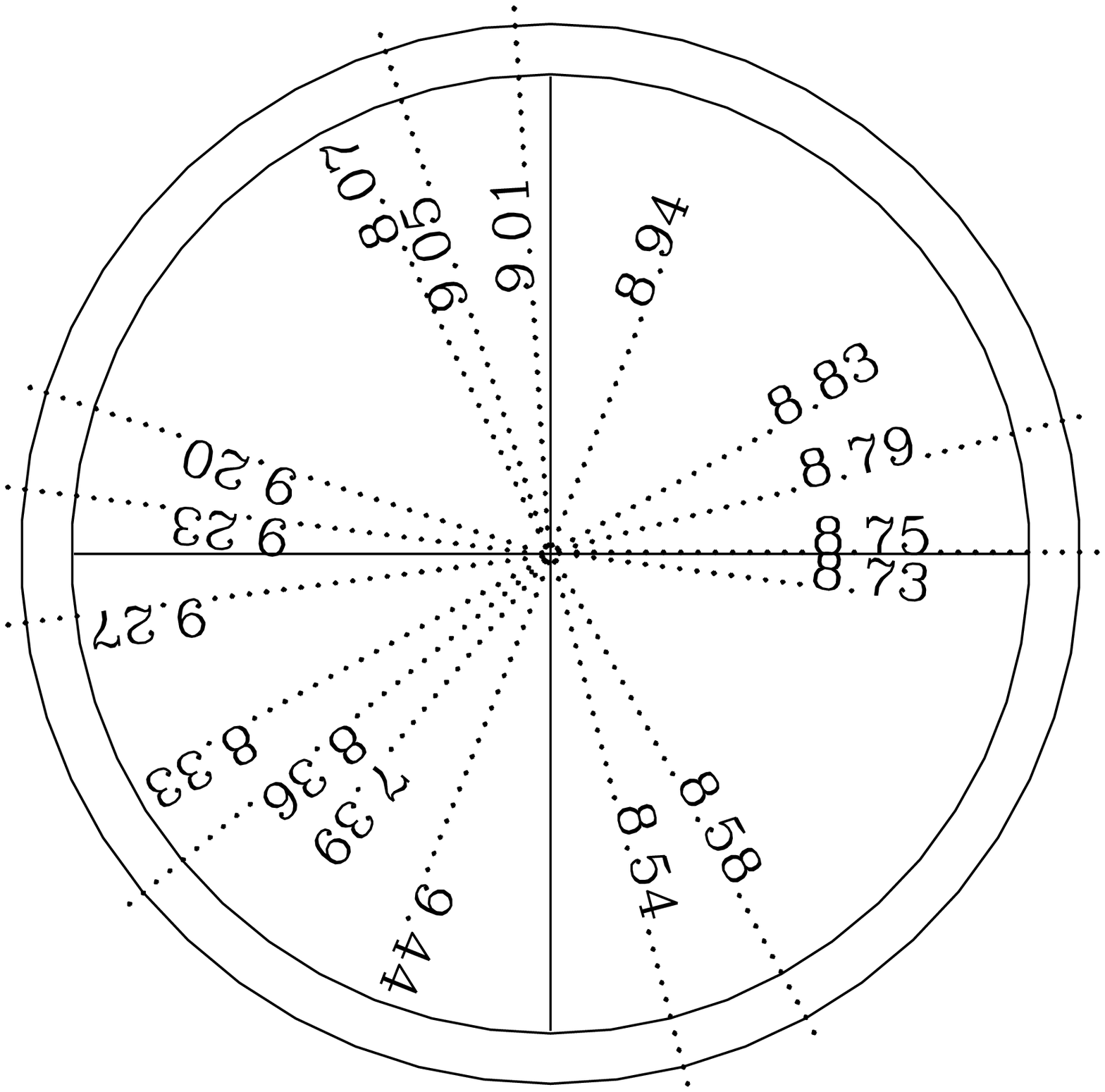}
\caption
{Top panel : A schematic representation of AG Peg system.
Bottom panel : The orientation of the binary system axis at 
different phases.}
\end{figure}

A very schematic representation of the system  and the angles of view at the
 different phases are given in Fig. 3 (top panel and bottom panel, respectively).

At $\phi$ = 8.07 we see almost all of  the shock fronts,  internal and external to the system 
(Fig. 3).
The distance of the reverse shock from the white dwarf,
r$_1$, is of the order of  the binary
separation (in Paper I it is found that the front is very close to
the  red giant surface), so, we adopt  r$_1$ =5 10$^{13}$ cm ;  the radius of the RD
expanding front  is r$_2$($\phi$=8.07)= r$_2$($\phi$=7.05)+$\Delta$ r, where
 $\Delta$ r is the distance  covered by the shock front between phases 8.07 and 7.05.
Then, r$_2$ = 3.2 10$^{15}$ cm (Paper I) + 7 10$^{14}$ cm (Sect. 4) is  $\sim$  4 10$^{15}$ cm.

Actually, (r$_2$/r$_1$)$^2$  $\sim$ 6400 at $\phi$ = 8.07 is  lower than 
w(m2)/w(m1) =  1.2 10$^4$ by a factor $<$ 2.
The ratios of the weights which appear in Table 2   are determined by the best fit of the calculated
 to  the observed data. The discrepancy by a factor of $<$ 2 indicates that the radiation  
emitted by the
reverse shock is slightly absorbed by the outer atmosphere of the red giant.
 At later phases, $\phi$ = 8.33 and $\phi$ = 9.27, the radiation  intersects  the red giant at angles 
that  depend on the phase angle, and the
path  length throughout the red giant, l  = 2 r$_{RG}$ cos $\alpha$ (r$_{RG}$ is the radius 
of the red giant and $\alpha$
is the angle between the binary system axis and the viewing direction from Earth), 
is higher at $\phi$=8.33 than at $\phi$=9.27.
 So, at $\phi$=9.27 the radiation
from the  reverse shock is less absorbed than at $\phi$=8.33, in agreement with the trend
of the w(m2)/w(m1) ratio  that  is 10$^5$ at $\phi$=9.27
 and   is 1.5 10$^5$  at $\phi$=8.33 .
The trend  is clearly depending on the phase angle,  however,  the trend   does not
depend only on the angle  because the radius of the expanding shock is not symmetric
around the system (see Nussbaumer 2000, Fig. 1).

Regarding the SD shock,   the ratio of the relative weights
w(m3)/w(m2)  does not follow any specific trend, because it depends on fragmentation 
of matter and on the filling factor.
 Also those ratios were chosen  by the  best fit to  the data.

Considering that the AG Peg system is very complex, we have focused on the most
significant lines, e.g. NV 1240, CIV 1550, HeII 1640, CIII] 1908, CII] 2325, etc. in the UV,  
and HeII 4686 in the optical range. 
 NIV] 1486 is overestimated by the calculations.
OI 1305, HeI 4471 and MgII are underestimated by a large factor.
The contribution from a model corresponding to a large low temperature region
could improve this result.
[NeV] 3426/ CIV  line ratio is overpredicted.  Notice, however, that this ratio refers to lines
emitted from one or two different nebulae, indicating that  more model components
should be considered. The same is valid for the [FeVII] 6087 /CIV line ratio. In this case, however,
 some iron could be locked into dust grains.

A slightly lower N/H could improve the fit of the NV 1240 and NIV] 1486 line ratios to CIV, 
particularly at phase  9.44.
 N/H approaching  to the ISM value,  and the relatively low \n0
and \B0 adopted for models m2 and m3, confirm that the merging process is  taking place.

In the averaged spectra we found \Ha/\Hb $<$ 3, lower than the observed line ratios, because the  weights
of m1 and m2 models are relatively high. A possible explaination is given by Shore (1992) who claims that 
in a moving medium with an internal velocity gradient, the optical depth of the line
 $\Delta$$\tau$ $\sim$ $\Delta$v/(dv/dl). $\Delta$v=$\Delta$$\nu_D$, where $\Delta$$\nu_D$ is the Doppler width
of a line with rest frequency $\nu_0$. The reduction of the velocity gradient causes a rapid 
increase in the line opacity , even if the density has dropped enough as to  make the equivalent static line profile
completely optically thin. Indeed, a velocity gradient is predicted in the outer atmosphere
of the red giant. Adopting \Vs=150 \kms at $\phi$ = 7.12 (Contini 1997), \Vs =100 \kms at the latest epochs,
and dl=10$^{14}$ cm, $\Delta$$\tau$ should increase by a factor $>$ 10$^7$, leading to higher \Ha/\Hb ~(see Sect. 2).

We can now compare  the luminosities of the lines  calculated by the models with   those observed by KPK01.
Considering, for instance [NeV] 3426 at $\phi$ = 8.07, the
 calculated  luminosity is $\sim$  3  10$^{30}$ erg s$^{-1}$  while the observed one (KPK01, Table 4) 
is 4 10$^{31}$ erg s$^{-1}$ adopting a distance to Earth, d=800 pc.
This line is mainly emitted from the nebula corresponding to model m1.
A better fit to the
observed luminosity could be obtained adopting \n0=3 10$^6$ \cm3 instead of 10$^6$ \cm3 for  m1.
The preshock density  used at $\phi$ = 7.05 (Paper I, Table 4)  was 3 10$^6$ \cm3.

The distance of the SD expanding nebula from the center of the binary system,
r$_3$, at $\phi$=9.44 is obtained
comparing the luminosities of  calculated and observed CII] lines.
From
CII]$_{obs}$ $\times$ d$^2$ = CII]$_{m3}$ $\times$  r$_3^2$,  r$_3$ results $\sim$ 10$^{16}$ cm.
The CII] line is considered  because it is emitted mainly from the outer SD nebula at photometric minimum.
The distance  is larger by a factor $>$ 2 than the distance of the
RD expanding  nebula corresponding to model m2. Even if the RD and SD expanding shocks are at
different distances from the system center, they may belong to the same wave. In fact, the "fingers"
created by R-T and K-H instabilities at the shock front can be elongated up to a large
fraction of the shock front radius (see Contini \& Formiggini 2001).

In summary,  the interpretation of the data
 by a  relatively small number of models is sound, even if approximated.
Anyway, some interesting results follow.

\subsection{Relative contribution of the different nebulae to the spectra}

\begin{table*}
\centerline{Table 3}
\centerline{Relative  contributions (in \%) to the line intensities}    
\small{
\begin{tabular}{ lllllllllllll l l l } \\ \hline\\
 & \multicolumn{3}{c}{$\phi$=8.07}&
 \multicolumn{3}{c}{$\phi$=8.33}&
 \multicolumn{3}{c}{$\phi$=9.27}&
 \multicolumn{3}{c}{$\phi$=9.44}\\

\ line & m1  & m2  & m3  & m1  & m2  & m3  & m1  & m2  & m3 & m1  & m2  & m3  \\ \hline\\

\ NV 1240     &  84.7 & 14.78   &   0.54& 30.11 & 68.75 & 1.13&37.60 & 61.38 & 1.02& 0.0 & 99.0& 0.99      \\
\ SiIV] 1394  & 78.90 & 3.68   &   17.40 & 34.4 & 20.98 & 44.61 & 44.66 & 17.7  & 37.64 & 0.0 & 43.69 & 56.31 \\
\ SiIV,OIV] 1403 & 91.35 & 5.39  & 3.25& 50.45 & 38.96& 10.59&61.04& 30.64 & 8.32& 0.0 & 85.86 & 14.14 \\
\ NIV] 1486  &  92.9   &   5.25 &  1.81  &  53.94  & 39.88 & 6.19 & 64.3  & 30.9 & 4.79& 0.0 & 91.41 & 8.59 \\
\ CIV 1550   &  92.27  &   3.57   &   4.16  &56.47 & 28.56 & 14.97 & 66.62 & 21.9  & 11.48& 0.0 & 75.89 & 24.11 \\
\ HeII 1640  & 86.94   &  13.05   & 0.014   &33.74 & 66.23 & 0.032&43.92 & 56.05 & 0.026 & 0.0 & 99.97 & 0.029 \\
\ OIII] 1664  & 67.78 &   1.35     &  30.87 & 25.39& 6.60 & 68.01 & 34.36 & 5.81 & 59.83& 0.0 & 13.81 & 86.19 \\
\ NIV 1719 &   79.63& 20.36 & 0.003&23.02& 76.98 & 0.0051& 31.51 & 68.49 & 0.0045&0.0 & 100. & 0.0 \\
\ NIII] 1750 &  64.84 & 1.35 & 33.8  & 23.04 & 6.29 & 70.66 & 31.54 & 5.60 & 62.86&0.0 & 12.81 & 87.19  \\
\ SiIII] 1892 & 30.21 & 0.64 & 69.14& 6.78& 1.89 & 91.32 & 10.0 & 1.82 & 88.11& 0.0 & 3.30 &96.70   \\
\ CIII] 1908 & 47.39 & 0.81 & 51.74& 13.07 & 2.93 & 84.0 & 18.78 & 2.74 & 78.48& 0.0 & 5.44 & 94.55 \\
\ CII] 2325 &  0.27 & 0.002 & 99.73 & 0.05 & 0.004  &99.95& 0.071 & 0.004& 99.93& 0.0 & 0.0074& 99.99 \\
\ [MgV] 2783 & 81.05& 18.94 & 0.0035& 24.65 & 75.34 & 0.006 &33.48 & 66.50 & 0.0056& 0.0 & 99.99& 0.0052 \\
\ MgII 2793 &  4.09 & 0.033 & 94.88& 0.90 & 0.077 & 99.02&1.38 & 0.077 & 98.54 & 0.0 & 0.128& 99.87\\
\ [NeV] 2976 &  85.60 & 14.37 & 0.032& 31.28 & 68.65 & 0.069& 41.19 & 58.75 & 0.058& 0.0 & 99.94 & 0.06\\
\ HeII 3203 & 87.0  & 12.98 & 0.013& 33.88 & 66.09 & 0.031& 44.08 & 55.89 & 0.026 & 0.0 & 99.97 & 0.028 \\
\ [NeV] 3426 & 73.3  & 26.66 & 0.0043 & 17.38 & 82.61 & 0.006 & 24.46 & 75.54 & 0.0056 & 0.0 & 100. & 0.0\\
\ HI 4340 &  86.7 & 12.93 & 0.37 & 33.61 & 65.56 & 0.836& 43.78 & 55.5  & 0.71 & 0.0 & 99.23 & 0.77 \\
\ HeI 4471 & 69.80 & 2.99 &27.2 & 25.95 & 14.56 & 59.49 & 35.03 & 12.77 & 52.2  & 0.0 & 28.76 & 71.24 \\
\ HeII 4686 &  85.94 & 14.09 & 0.009 & 31.78 & 68.20 & 0.012 & 41.75 & 58.23 & 0.017 & 0.0 & 99.98 & 0.0177\\
\ \Hb ~4861   & 100. & 0.000  & 0.0 &100.0 & 0.0   & 0.0 & 100.0 & 0.0   & 0.0 & 0.0 &99.99 & 0.007\\
\ [FeVII] 6087 & 72.84 & 27.16& 0.0015 & 17.02 & 82.98 & 0.002  & 23.98 & 76.0  & 0.002& 0.0 & 100. & 0.0 \\
\ \Ha ~6563 & 85.77 & 12.79 & 1.43& 32.8  & 63.98 & 3.22 & 42.89 & 54.38 & 2.74 & 0.0& 97.04 & 2.96\\
\hline\\
\end{tabular}}
\end{table*}

The relative contribution  of the  emitting nebulae to  
each line is shown in Table 3.
The weight of  models m2 and m3, representing the  nebulae propagating outwards
in the atmosphere of the red giant,  relative to the weight of the nebula corresponding to the reverse
shock between the stars increases with the angle
determined by the phase fraction (.07, .27, .33, and .44), 
indicating that the  external  nebula is not  present around the whole system,
but  only in the region opposite to the WD.

The nebula downstream of the reverse shock dominates at $\phi$ = 8.07 
except for  a few lines (SiIII], CIII], CII], and MgII). A large contribution to
these lines comes from the SD expanding shock. The RD expanding  shock
 contributes partly to NIV, [NeV], and to [FeVII].

The relative contribution of the three nebulae to the lines
are  similar at $\phi$ = 8.33 and at $\phi$ = 9.27, 
the reverse shock being less dominant and indicating that the relative contributions
of the nebulae depend on the viewing angle
of the system rather than on the phase. At both phases 8.33 and 9.27
the RD expanding shock is dominating for  most of the lines (NV, HeII 1640, NIV, [MgV],
[NeV], HeII 3203, [NeV] 3426, HI 4340, HeII 4686, [FeVII], and \Ha). 
The SD  expanding shock dominates the low ionisation level lines.

At $\phi$ = 9.44 the contribution from the reverse shock becomes negligible
because  the nebula between the stars is  occulted by the red giant.
The high ionisation level lines are produced mainly by the RD expanding shock
and low ionisation level  lines  come from the SD nebula.

 The relative contribution of the  nebula  between the stars to the HeII 1640 line
decreases monotonically with  phase  fractions .07, .27, .33, and .44
On the other hand, the relative contributions of
m2 to NV 1240 are 14.8\%, 61.4\%, 68.75\%, and 99.0\%, at the same angles, respectively.

Concluding,  the evolution of the spectra is  more likely related 
 to the viewing angle of the system, not particularly  to the phase
between $\phi$ = 8.07 and $\phi$ = 9.44.
 We do not  have data  relative to viewing angles higher than 0.44, which
could  hopefully strengthen  the results.

\subsection{Evolution of the line fluxes with phase}

It is now possible to understand the trend of the lines shown in Fig. 1.
Table 3 shows that the contribution from the SD shocked nebula (m3) to  \Ha  ~is  only a few \% ,
therefore, the calculated \Ha/\Hb ~line ratio  is likely to be $\sim$ 3
throughout the whole period between phase 8.07 and 9.44.
On the other hand, the observed value is  $\geq$ 5 towards photometric minimum.
This high value may be explained considering that  the RD expanding  shock 
is represented by a composite model
which accounts for both the photoionisation by the WD and the shock.
The side of the emitting nebula facing the observer at  photometric minimum is  shock dominated
(Fig. 2, middle diagram).
So, in this phase we see mainly  \Ha ~and \Hb ~fluxes from this side of the nebula, 
which  are  produced by collisional ionisation. Moreover, the optical
depth in the radiation dominated region is high (see Sect. 4.1) leading to
high  \Ha/\Hb . 

HeII 4686 and [FeVII] do not show a periodic change with phase
because a large fraction of these lines is emitted by the nebula downstream of the RD expanding  shock.
So, the decrease of the reverse shock contribution at phases approaching
the photometric minimum  is compensated by the large contribution from the RD expanding shock.
These lines show, therefore, less impressive variability up to $\phi$ $\geq$ 9.2.
At $\phi$ = 9.27 the calculated F$_{\lambda 4686}$ is  0.11 \erg, while at
$\phi$ = 9.44 F$_{\lambda 4686}$ = 0.06 \erg. The  decrease of the calculated flux
between these two phases  is  in agreement with the observed decrease shown in Fig. 1
after $\phi$ = 9.27.

HeI 4471  is emitted
mainly by the SD expanding shock (m3) and the flux is  very low, F$_{\lambda 4471}$= 2 10$^{-5}$ \erg.
Therefore,  this line flux shows a maximum in Fig. 1
when the system is  near maximum. 
 The same is valid for H$\alpha$ and \Hb.
Notice, however that  HeI 4471/CIV line ratio is underpredicted by the models.

All the lines show an asymmetric behaviour before and after maximum. This is explained by
the fact that the emitting nebulae are not  symmetric relative
to the axis joining the center of the WD with the center of the red giant.
In the  hydrodynamical picture of symbiotic systems  presented by Nussbaumer (2000, Fig. 1 left) 
the wake is, in fact, strongly asymmetric, so that the emitting nebulae 
could  partly disappear in some  location at angles corresponding to high phase fractions.

\section{Concluding remarks}

The analysis of the spectra from the HST and from the FAST spectrograph at the Whipple Observatory
reported by Kenyon et al (2001) at different phases leads to
new  results about the configuration of AG Peg.
The FAST spectra contain both H$\alpha$ and \Hb ~lines,  providing the tool for
modelling the system at different phases. 
Interestingly, the  trend of the periodical variability of \Ha/\Hb , as well as the trend of
the most significant lines  indicate that 
different nebulae contribute to the final spectra.
In particular, the high \Ha/\Hb ~ line ratios ($\sim$ 5) suggest  that shock dominated models
must be accounted for. Moreover, the trend of the HeII 4686 line variability with phase
indicates that radiation dominated nebulae are  strongly contributing  to the spectra also near the
photometric minimum.

Consistent modelling by the code SUMA, which accounts for the coupled effect of photoionisation
from the hot star and  shocks, shows that at least three nebulae provide the
spectral components :  1) the nebula between the stars, downstream of the  shock created by head-on 
collision of the winds, 
which propagates in reverse towards the WD, 2) the nebula downstream of the head-on-back  shock 
expanding in the outskirts of the red giant atmosphere  and reached by the photoionising flux from the WD, 
and 3)  the nebula downstream of the  same shock front which is shock dominated, i.e. it is not
reached by the photoionising radiation flux.
In fact, the matter in this region is very clumpy because the shock front
is disrupted by R-T instability, and radiation  from the WD is
prevented  from reaching the region by the intervening  clumps in some locations. 

The fit of calculated to observed line ratios is acceptable enough to justify
the choice of the models,
so that the  relative contribution of the different nebulae to the final spectra at
different phases  can be calculated for all the lines.
It is found that the contribution of the  nebula between the stars  prevails for most of the lines
near maximum ($\phi$ = 8.07), declining towards higher phases and even disappearing at phase 9.44.
Low ionisation level lines (e.g.  CII] and MgII)   are mainly emitted from the SD expanding nebula.
The evolution of the spectra between phases 7.34 and 9.44 is  more likely related 
with the viewing angle of the system, and not to the temporal evolution of the system.

Models m2 and m3 which represent the expanding shock are characterized by a relatively
low preshock density and magnetic field, much lower than found by modelling
AG Peg at $\phi$ $\leq$ 7.12 (Paper I). This suggests that the shock,  expanding in the red
giant atmosphere opposite to the WD, has already reached the outskirts of the system and is 
merging with the ISM. At $\phi$ = 9.44 the expanding shock has swept up
a large portion of  matter with relatively low density. The line and continuum fluxes 
emitted from downstream the shock are $\propto$ n$^2$ and, therefore,  decrease  
as more IS matter is swept up,  
leading to the slow decline of AG Peg  continuum and line fluxes observed  by KPK01.

\section*{Aknowledgments}

I am very grateful to the referee for many enlightening suggestions
and precious remarks,
and to D. Prialnik and S. Beck for reading the manuscript.

\bsp

\label{lastpage}

\end{document}